# Properties of a Stochastic Model for Life Table Data: Exploring Life Expectancy Limits

## Christos H Skiadas[1] and Charilaos Skiadas[2]

**Abstract:** In this paper we explore the life expectancy limits by based on the stochastic modeling of mortality and applying the first exit or hitting time theory of a stochastic process. The main assumption is that the health state or the "vitality", according to Strehler and Mildvan, of an individual is a stochastic variable and thus it was introduced and applied a first exit time density function to mortality data. The model is used to estimate the development of mortality rates in the late stages of the human life span, to make better fitting to population mortality data including the infant mortality, to compare it with the classical Gompertz curve, and to make comparisons between the Carey med-fly data and the population mortality data estimating the health state or "vitality" functions. Furthermore, we apply the model to the life table data of Italy, France, USA, Canada, Sweden, Norway and Japan, and we analyze the characteristic parameters of the model and make forecasts.

## Introduction

The Health State of human beings is a term generally accepted as a crucial factor of the human life and is connected, even not verified quantitatively, to the life expectancy. The human health can be considered as a stochastic variable as it is strongly associated with uncertainties having to do with various factors both from the environment but also from the internal mechanism and information included into the DNA and the genes. The probability of sudden changes in the state of the human health due to illnesses or accidents is quite large, strengthening the assumption that the state of the human health could be considered as a stochastic variable. It is known that human characteristics vary around mean values, and the deviations seem to be the physical consequence of the probabilistic-stochastic character of the process. Moreover the probabilistic or better the stochastic character of the health state of individuals is generally accepted and measured at least qualitatively. Attempts to quantify the process go back to Strehler and Mildvan (1960) (1) who suggest the term "vitality" of a person, a stochastic function with a decreasing drift and changing randomly with age during the human lifetime. The end comes when the stochastic path crosses for the first time the zero line, representing the zero level of vitality or the zero health state.

## The Stochastic Modeling

---
[1] Technical University of Crete, Data analysis and forecasting laboratory, Chania, Crete, Greece.
E-mail: skiadas@asmda.net
[2] Department of Mathematics and Computer Science, Hanover College, Indiana, USA.
E-mail: skiadas@hanover.edu

*Paper version: January 10, 2011*

Based on a stochastic process for the health state function, we had explored the problem of modeling human mortality in 1995, when we developed and applied a dynamic model based on the first exit time or hitting time theory. That model is flexible enough as to account for expressing the human mortality distribution at all ages, including infant mortality (see the paper by Janssen and Skiadas, 1995 (2)). The proposed first exit or hitting time probability density function $g(t)$ includes 5 parameters. The two extra parameters account for the infant mortality, making it evident that a satisfactory model with fewer parameters could be obtained.

More recently (3, 4) we have developed more flexible stochastic models to express the human mortality. Though it is impossible to find the stochastic path for the state of health for every individual, it is possible to find the Mean Value of the health state or vitality of a population as a summation of the related health states of every individual. Starting with a model for the mean health state function $H(t)$ we can proceed in the construction of a stochastic model for the state of the human health, solve it, find the related probability density function $p(t)$ and then estimate the first hitting time probability density function $g(t)$ from a barrier located at zero for the stochastic process expressing the state of the human health. The parameters of the last function $g(t)$ are estimated by fitting it to the data collected from the Bureau of the Census for the yearly deaths of the population in a country.

Our approach was mainly to find models with explanatory power and able to fit the data directly instead of applying models to transformed data via logarithms as is the case in several applications starting from Gompertz (1825) (5). It was evident that the methods for nonlinear regression where not well developed when Gompertz introduced his model and the linearization via a logarithmic transformation was a reasonable substitute when working with mortality data. However, we should notice that by taking logarithms of the data and then fitting, the results highly underestimate the error term. The real error is of the order of the exponent of that estimated when taking logarithms.

Few years after the paper by Janssen and Skiadas (1995), Weitz and Fraser (2001) (6) proposed the inverse Gaussian as a model expressing the first hitting time probability density function for modeling the life time process of the med-fly based on the data collected and presented by Carey in a publication in Science (1992). The model proposed by Weitz and Fraser is nothing else but the classical model proposed by Schrödinger (1915) (7) and Smolukovsky (1915) (8) in the same journal issue. This model suggests a linear function for the mean value of the stochastic process thus assuming a decay process for the mean value of the health state. This is a correct argument in the case modeled by Schrödinger and Smolukovsky as they considered the spontaneous emission of particles from a radioactive material, but it is not appropriate for modeling the state of the human health. According to our experience the mean value of the human health could be stable or even increasing at the young ages and then decreasing slowly at the mid ages and fast decreasing at the old age. Even the simplest function should be non-linear. This is the underlying assumption in the theory proposed by Strehler and Mildvan in the 1960 publication in Science (1).

**Properties of the Stochastic Model**

However, the Weitz and Fraser approach gave good results in the case of the med-fly data, only differing from the nonlinear mean value by a small curvature as is presented in the paper by Skiadas and Skiadas (9). The data on med-flies are collected from an experiment (Carey et al. 1992 (10)) in which the life duration of 1.203.646 med-flies in cages was measured. As it was presented in the same comparative study by Skiadas and Skiadas (9) the inverse Gaussian is a right skewed model appropriate for some modeling approaches but not for the highly left skewed human life table data. The figure 1A for the mortality data for USA 2004 for females compared with the Carey med-fly data is evident. The age axis is rescaled as to account for both human and med-fly data. The rescaling is according to the method proposed by Robine and Ritchie (1993) (11).

The first exit time model used is of the following form (3)

$$g(t) = \frac{k(l+(c-1)(bt)^c)}{\sqrt{t^3}} e^{-\frac{(l-(bt)^c)^2}{2t}}$$

Where $b$, $l$, $k$ and $c$ are parameters. The key points in deriving this model where: 1) the proposal of a stochastic differential equation for the health state or the vitality (according to Strehler and Mildvan) of an individual, 2) the derivation of the transition probability density function $p(t)$ associated with the stochastic variable for the health state (vitality) by solving the appropriate Fokker-Planck partial differential equation and 3) the derivation of the first exit time probability density function $g(t)$ from the stochastic process having transition probability density function $p(t)$. These three steps where first achieved by Janssen and Skiadas (1995) and later Skiadas and Skiadas (10) suggested the model presented by the above hitting time density function. This model is already applied in mortality data for more than 35 countries and is quite flexible as to cover the infant mortality as well. The data are obtained from the Human Mortality Database (www.mortality.org) and the related applications along with the fitting programs appear in the webpage http://www.cmsim.net/id14.html .

The function used for the health state is: $H(t)=l-(bt)^c$. When $c=1$ the simple case studied by Schrödinger (1915) and Smolukovsky (1915) and applied by Weitz and Fraser (2001) results. As it is obvious from Figure 1A the Carey med-fly data is right skewed whereas the human mortality data, as seen in the case of USA females in 2004, is highly left skewed. The model is applied to the data with the estimates $b=0.2479$, $l=8.948$, $k=0.3803$ and $c=1.3$, $R^2=0.992$ for Carey med-fly data and $b=0.0194$, $l=12.705$, $k=0.3677$ and $c=5.3$, $R^2=0.988$ for females in USA 2004.

A very interesting case appears when we draw the tangent lines passing from the left inflection points of the curves fitted to med-fly and female (USA 2004) mortality data. Both lines, illustrated in Figure 1A, are almost parallel. The estimated values of the first derivative of the first exit time density functions are 0.00301 for the Carey med-fly data and 0.00327 for the USA 2004 female data, slightly deferring between each other. It is obvious that at least for this example the death rates at high ages are similar. Instead the behavior is different in the tails as the med-fly data present a very long tail compared to the USA 2004 female data.

The representation for the health state functions is given in the related graph in Figure 1B. The health state function is $H(t)=l-(bt)^c$. The data for males in USA 2004 are also fitted. The estimated parameters are $b=0.0218$, $l=10.936$, $k=0.362$ and $c=4.6$, $R^2=0.964$. The higher level accounts for the females in USA compared to USA males. The model suggests an almost linear development for the first decades of the human life span, a gradual decrease in the mid ages, and finally a fast decrease at old age. Instead for the Carey med-fly an almost linear decrease of the health state is estimated. The curvature expressed by the exponent $c=1.3$ slightly differs from the linear one ($c=1$) suggested by Weis and Fraiser. Instead for the human health states the exponent is $c$=5.3 and 4.6 for females and males in USA respectively, suggesting a fairly large curvature.

A very important point is to explore the properties of the Health State or Vitality function based on the curvature. It is clear that the curvature is larger as the exponent $c$ takes higher values. Larger curvature is associated with a life span in the area of high vitality expressed by the first and almost linear part of the health state curves (see Figure 1B for males and females in USA). In the following we can estimate the point where the curvature $K(t)$ expressed by the following formula takes its maximum value.

$$K(t) = \frac{|c(c-1)b^c t^{c-2}|}{(1+c^2 b^{2c} t^{2c-2})^{3/2}}$$

This maximum value for the curvature could be connected with the maximum deterioration of the physical repair mechanisms of the human organism (12). Estimating the maximum for the curvature we find that it is obtained at 77 years for the females and 74 years for males in USA the year 2004. Instead for the med-fly data the maximum curvature is obtained in the first year of the life span. We can assume that the mechanisms underlying the vitality of the med-fly are totally different from the related mechanisms of the human beings. Another interesting point resulting from the study of the curvature of the health state function is that it is smaller and smaller as we are moving at higher ages from its maximum value. This is in accordance with the assumption of slower mortality rates at higher ages (13, 14).

In the following we will examine how the mortality evolves at the late stages of the life span by using the first exit time density function. This will be checked by a comparative study (Sweden, females). The issue can be explored by observing the tangent lines at the right inflection point of the regression curve. This area accounts for the late years of the human life span that could be noted as the longevity area. We have drawn these lines for the case of females in Sweden from 1800 to 2000 illustrated in Figure 1C. It is clear that the tangent line is moving to the right part of the figure to the higher ages. This is in accordance with the improvement of the health state during the 200 years elapsed from 1800 to 2000. However, one should expect parallel lines moving to the right due to the significant improvement of the living conditions in our societies. To our surprise the tangent lines, while moving to the right, turn vertically and tend to consistently meet the horizontal axis just few a years after 100. The right part of the first exit time density function $g(t)$ is mainly expressed by the tangent line while a small part accounts for people included in the tail of the density function. The expectations for a longer life are mainly connected to the way the tangent line is moving to the right to the higher ages. The observations of the last decades show that the tangent line tends to gradually stabilize in a vertical position than to move significantly to the right. This can be due to the way our repair mechanisms included in the DNA and genes decide for the life expectancy. If it is so, the tangent line will be more and more vertical and the form of the first exit time density function at the maximum area more sharp following improvements in the living standards. Instead, if our destiny is not driven by the code, the tangent lines should move parallel to the right. The later may result following the improvement of biological sciences. However, in the following, we make parameter analysis of the mortality data for various countries to search for the future of human longevity.

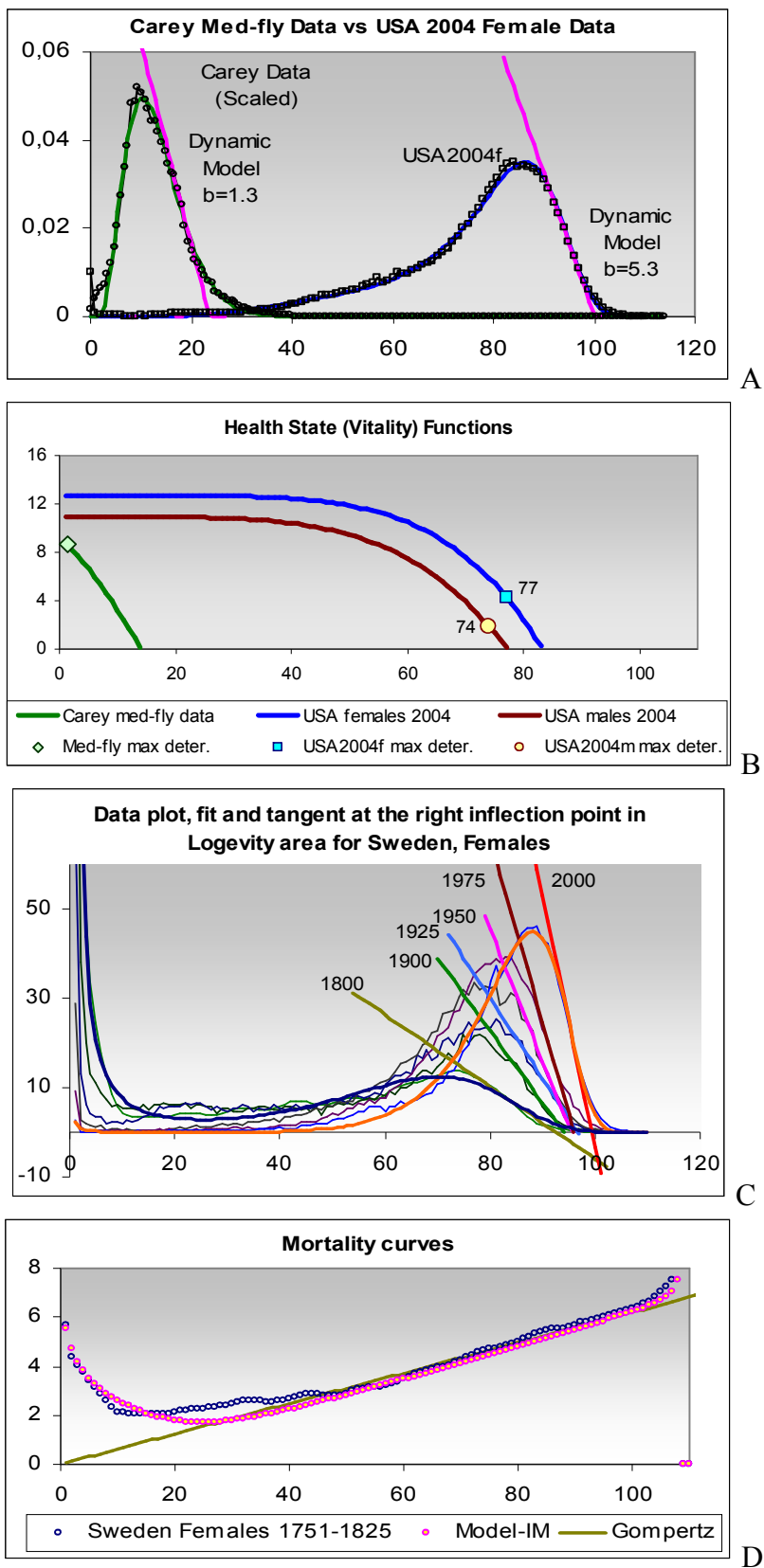

Figure 1. Applications of the First Exit Time Model.
A. Comparison of Carey med-fly data with the USA data for females, 2004.
B. Form of the Health State Functions for Carey data and USA data
C. The characteristic tangent for the Sweden data for females
D. The mortality curves

As we have proposed and applied a model presenting a good fit to the data, including even the infant mortality we will now try an application to the data that perhaps Gompertz was aware at the time (1825) when he published his famous paper, and compare it with what Gompertz could see using his formula. The data used is the mean value of the females' deaths in Sweden from 1751 to 1825. The resulting mortality curves are illustrated in Figure 1D. As it was expected our model expresses the infant mortality as well, whereas the Gompertz model is expressed by the characteristic line. The data are transformed to account for 1000 people and the logarithm of the death rate is computed for every year of age. The estimated parameters of the first exit time model are $b=0.0223$, $l=0.5878$, $k=0.552$ and $c=4.3$ and for the Gompertz line $b=0.062$.

**Parameter Analysis**

The parameters of the first exit time model proposed are estimated for several countries and for females for a quite large time period. The results are illustrated in Figure 2. Seven countries are selected: Italy, Sweden, USA, Japan, Canada, France and Norway. The parameters are estimated for as many larger periods as possible according to the data are provided from the Human Mortality Database. In Figure 2A the parameter $b$ is illustrated. The parameter values after 1950 form a group of almost parallel and slowly decreasing lines. In Figure 2B the parameter $c$ is presented. The exponent $c$ is growing in all the cases with Sweden, Italy, France and Norway to reach higher values. In figure 2C the parameter $l$ is presented. This parameter accounts for the infant mortality data period and, as the infant mortality tends to decrease, the parameter $l$ tends to zero. In this case we will have a simpler model of the form:

$$g(t) = \frac{k(c-1)(bt)^c}{\sqrt{t^3}} e^{-\frac{(bt)^{2c}}{2t}}$$

As the parameter $k$ is a renormalization constant, the process is characterized by the parameters $b$ and $c$. An illustrative example is given in Figure 2D. Two possible scenarios resulting from the projection of the parameters $b$ and $c$ to 2050 are presented: $b=0.0125$, $c=14$ and $b=0.0115$, $c=14$ are selected and the population is equal to 1000.

The maximum death rate is achieved when: $T_{max} = \left[\frac{2c-3}{2c-1}\right]^{1/(2c-1)} \frac{1}{b^{2c/(2c-1)}}$

For relatively large $c$ an approximation is of the form: $T_{max} \approx \frac{1}{b^{2c/(2c-1)}}$

For the previous example the estimates are $T=93.82$ years for $b=0.0125$ and $c=14$ and $T=102.3$ years $b=0.0115$ and $c=14$.

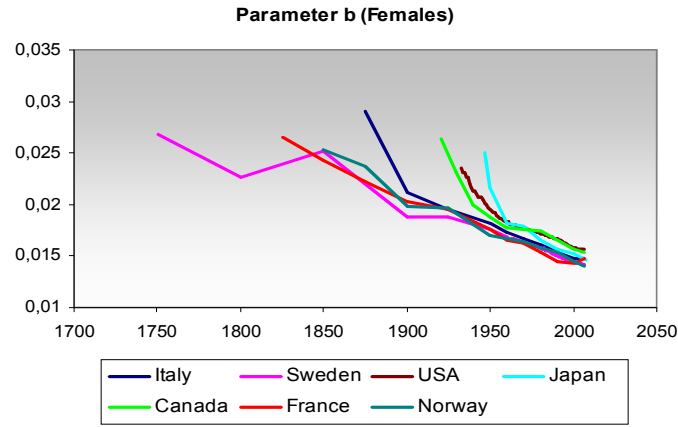

A

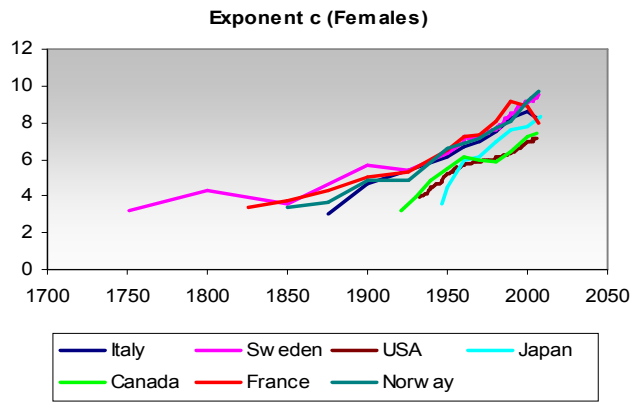

B

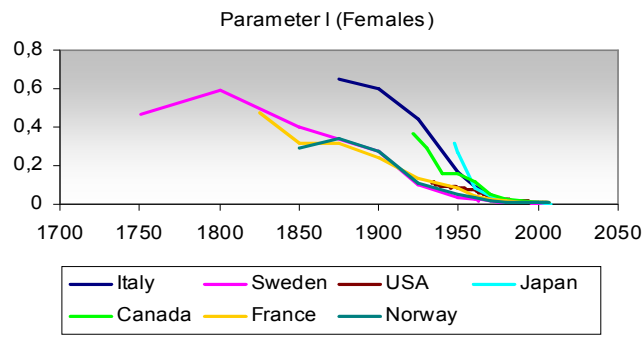

C

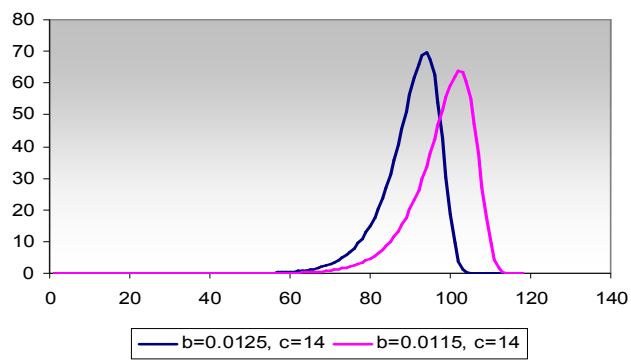

D

Figure 2. Parameter's values for various countries

**Summary and Conclusions**

In our studies on modeling and explaining the mortality laws we have used the first exit or hitting time theory for a stochastic process and we have applied the related formulas to human life table data and we have done comparisons with data as the Carey med-fly data that where presented, analyzed and compared with human mortality data.

There is a controversy regarding existing limits in life span or if it is prescribed through the DNA code and genes, and the influence of external factors affecting longevity. Perhaps the answer will come later following the advancement of biosciences. Today we can use the tools developed during the last centuries both from the analytic and also the applied point of view. We have mortality data that is quite well selected and provided by the bureau of the census of every country, models to apply starting from the famous Gompertz model, and theories regarding the life span and fitting techniques to data.

It was evident from the very early approaches that the stochastic character of the human life duration was inherent and underlying of every type of model that we could formulate, but the mathematic-analytic tools where less developed. For several studies the stochastic model applied was the inverse Gaussian, a model best suited to express a simple decay process, like for instance radioactive emission or the aging process of a simple organism. The decay process for this model is of the linear form $H(t)=l-bt$ instead of a nonlinear form of the type $H(t)=l-(bt)^c$ that we have suggested. In the later case it was found that very important results arise from the study of the curvature of this function.

Another point is that, due to the strong influence of the Gompertz technique in linearizing the mortality data by taking logarithms for the mortality rate, a large number of studies is directed to analyze the tail of the probability density function and the small amount of the history included in this part of mortality data. Instead the information included in the non-transformed mortality data and especially in the right part of the non-symmetric bell-shape form of the mortality data graph is more essential to understand the results of the unknown aging mechanisms. In the absence of any systematic aging mechanism (internal or external) the longevity tangent is expected to show no-characteristic behavior. Instead by analyzing the females' mortality data from Sweden of last two centuries we observe that the longevity tangent turns clockwise quite systematically. A further parameter analysis suggests a probability of increase of the life span in the future according to some scenarios.

Now it looks that with our systems of life care, the maximum death rate in the short run will face a limit close to 100 years of age. If it is so the tangent line in the longevity area will be more and more vertical during the years and the form of the first exit time density function at the maximum area more sharp following the improvement of the living standards. Instead, if our destiny is not driven by an internal mechanism the tangent lines should move parallel to the right. It remains to the improvement of the DNA and gene science and the related scientific techniques to improve our maximum life span.

**Acknowledgments**